\documentclass[twocolumn,showpacs,floats,prb,aps,unsortedaddress,superscriptaddress]{revtex4-1}
\usepackage[dvips]{graphicx}
\usepackage{amsfonts}
\usepackage{amsmath}
\usepackage{amssymb}
\usepackage{exscale}
\usepackage{eufrak}
\usepackage{afterpage}
\usepackage{upgreek}
\usepackage{color}	
\usepackage{braket}	
\usepackage{multirow}	
\usepackage{subfigure}
\usepackage{epsfig}
\usepackage{ifpdf}
\usepackage{epstopdf}

\usepackage[normalem]{ulem}    

\begin{document}

\title{Tunable two-dimensional electron system at the (110) surface of SnO$_2$}

\author{J.~Dai}
\affiliation{Universit\'e Paris-Saclay, CNRS, CSNSM-Centre de Sciences Nucl\'eaire 
			et de Sciences de la Mati\`ere, 91405, Orsay, France}
			
\author{E.~Frantzeskakis}
\affiliation{Universit\'e Paris-Saclay, CNRS, CSNSM-Centre de Sciences Nucl\'eaire 
			et de Sciences de la Mati\`ere, 91405, Orsay, France}
\affiliation{Universit\'e Paris-Saclay, CNRS,  Institut des Sciences Mol\'eculaires d'Orsay, 
			91405, Orsay, France}
			
\author{F.~Fortuna}
\affiliation{Universit\'e Paris-Saclay, CNRS, CSNSM-Centre de Sciences Nucl\'eaire 
			et de Sciences de la Mati\`ere, 91405, Orsay, France}

\author{P.~L\"omker}
\affiliation{Peter Gr\"unberg Institut (PGI-6), Forschungszentrum J\"ulich GmbH, 
D-52428 J\"ulich, Germany}
\affiliation{Photon Science, Deutsches Elektronen-Synchrotron DESY, 
Notkestr. 85, 22607 Hamburg, Germany}

\author{R.~Yukawa}
\affiliation{Photon Factory, Institute of Materials Structure Science, 
High Energy Accelerator Research Organization (KEK), 1-1 Oho, Tsukuba 305-0801, Japan}

\author{M.~Thees}
\affiliation{Universit\'e Paris-Saclay, CNRS, CSNSM-Centre de Sciences Nucl\'eaire 
			et de Sciences de la Mati\`ere, 91405, Orsay, France}
\affiliation{Universit\'e Paris-Saclay, CNRS,  Institut des Sciences Mol\'eculaires d'Orsay, 
			91405, Orsay, France}

\author{S.~Sengupta}
\affiliation{Universit\'e Paris-Saclay, CNRS, CSNSM-Centre de Sciences Nucl\'eaire 
			et de Sciences de la Mati\`ere, 91405, Orsay, France}

\author{P.~Le F\`evre}
\affiliation{Synchrotron SOLEIL, L'Orme des Merisiers, Saint-Aubin-BP48, 91192 Gif-sur-Yvette, France}

\author{F.~Bertran}
\affiliation{Synchrotron SOLEIL, L'Orme des Merisiers, Saint-Aubin-BP48, 91192 Gif-sur-Yvette, France}

\author{J.~E.~Rault}
\affiliation{Synchrotron SOLEIL, L'Orme des Merisiers, Saint-Aubin-BP48, 91192 Gif-sur-Yvette, France}

\author{K.~Horiba}
\affiliation{Photon Factory, Institute of Materials Structure Science, 
High Energy Accelerator Research Organization (KEK), 1-1 Oho, Tsukuba 305-0801, Japan}

\author{M.~M\"uller}
\affiliation{Peter Gr\"unberg Institut (PGI-6), Forschungszentrum J\"ulich GmbH, 
D-52428 J\"ulich, Germany}
\affiliation{Fakult\"at Physik, Technische Universit\"at Dortmund, D-44221 Dortmund, Germany}

\author{H.~Kumigashira}
\affiliation{Photon Factory, Institute of Materials Structure Science, 
High Energy Accelerator Research Organization (KEK), 1-1 Oho, Tsukuba 305-0801, Japan}
\affiliation{Institute of Multidisciplinary Research for Advanced Materials (IMRAM), 
Tohoku University, Sendai 980-8577, Japan}

\author{A.~F.~Santander-Syro}
\email{andres.santander-syro@u-psud.fr}
\affiliation{Universit\'e Paris-Saclay, CNRS, CSNSM-Centre de Sciences Nucl\'eaire 
			et de Sciences de la Mati\`ere, 91405, Orsay, France}
\affiliation{Universit\'e Paris-Saclay, CNRS,  Institut des Sciences Mol\'eculaires d'Orsay, 
			91405, Orsay, France}

\date{\today}


\begin{abstract}
We report the observation of a two-dimensional electron system (2DES) 
at the $(110)$ surface of the transparent bulk insulator SnO$_2$, 
and the tunability of its carrier density by means of temperature or Eu deposition. 
The 2DES is insensitive to surface reconstructions and, surprisingly, it survives even after exposure 
to ambient conditions --an extraordinary fact recalling the well known catalytic properties SnO$_2$.
Our data show that surface oxygen vacancies are at the origin of such 2DES, 
providing key information about the long-debated origin of $n$-type conductivity in SnO$_2$, 
at the basis of a wide range of applications.
Furthermore, our study shows that the emergence of a 2DES in a given oxide 
depends on a delicate interplay between its crystal structure 
and the orbital character of its conduction band.
\end{abstract}
%
\maketitle

\section{Introduction}
Tin oxide, a transparent insulator, 
is a technologically important compound with a wide range of applications. 
Its exceptional properties stem from its ability to exhibit variable oxygen stoichiometries 
--a consequence of the variable valence of Sn-- accompanied by substantial changes,
up to two orders of magnitude, in its conductivity. 
This remarkable combination of reducibility and changes in conductivity is crucial 
in the fields of gas sensing~\cite{Batzill2007, Batzill2003, Merte2017, Pearse, Sinner2002, 
Semancik1987, Watson1984, Cox1989}
and heterogeneous catalysis~\cite{Merte2017, Batzill2005bis, 
Jones1997, Berry1981, Batzill2005, Harrison1999, Cox1989, Solymosi1976, Fuller1973}, 
while the unique association of high conductivity due to intrinsic defects, 
transparency to visible light, and high resistance to chemical attack at ambient conditions, 
allow novel applications including solar cells, liquid crystal displays, 
or even transparent electrodes, conductive coatings 
and windows~\cite{Batzill2005, Maki2001, Cox1989, Sinner2002, Egdell1999}.

The origin of $n$-type conductivity in SnO$_2$, at the heart of its numerous applications,
has thus attracted a large scientific interest. 
Bulk SnO$_2$ stabilizes in the rutile structure, 
with a band gap of 3.6~eV~\cite{haines1997x, reimann1998experimental, Robertson1979, Munnix1983}, 
but the presence of a surface modifies this simple picture. 
Calculations on the thermodynamically most stable termination, 
the $(110)$ surface, have proposed O $2p$-derived defect states 
within the band gap~\cite{Munnix1983}. 
Early photoemission studies on oxygen deficient SnO$_2$ 
confirmed the existence of such defect states 
near the valence band maximum~\cite{Cox1988, Egdell1986, Cox1982, Themlin1990, 
Cox1988, Egdell1982, Egdell1987, Hollamby1993},
and suggested the gradual filling of the band gap
upon further reduction~\cite{Oviedo2000, Cox1988, Jones1997, Marley1965}. 
This is in contrast with the metallic two-dimensional electron systems (2DESs) 
observed in similar transparent conducting oxides, such as In$_2$O$_3$~\cite{Zhang2013}, 
CdO~\cite{Piper2008}, ZnO~\cite{Rodel2018, Yukawa2016} and TiO$_2$~\cite{Rodel2015, Rodel2016},
where surface defects lead instead to band bending and $n$-type doping of the conduction band.

By means of Angle Resolved PhotoEmission Spectroscopy (ARPES), 
here we prove the existence 2DES at the $(110)$ surface of SnO$_2$,
and characterize its electronic structure. 
Similar to SrTiO$_3$~\cite{Santander2011, Rodel2014} 
and KTaO$_3$~\cite{Bareille2014}, 
the 2DES in SnO$_2$ arises from oxygen vacancies, 
and is robust with respect to various surface reconstructions.
Intriguingly, the 2DES in SnO$_2$ is not affected by exposure to ambient air,
and can be observed without particular surface treatment.
Moreover, its carrier density can be tuned by temperature or deposition of
Al or Eu, which create oxygen vacancies at the SnO$_2$ surface,
in analogy to previous observations in other oxides~\cite{Rodel2016, Lomker2017, Frantzeskakis2017}.
These results open possibilities of using and controlling 
the surface conductivity of SnO$_2$ for novel technological applications.

\section{Materials and Methods}

\subsection{Crystal structure and Brillouin zone of $\textrm{SnO}_2$}
\begin{figure}[h] 
   \centering
   \includegraphics[clip, width=0.5\textwidth]{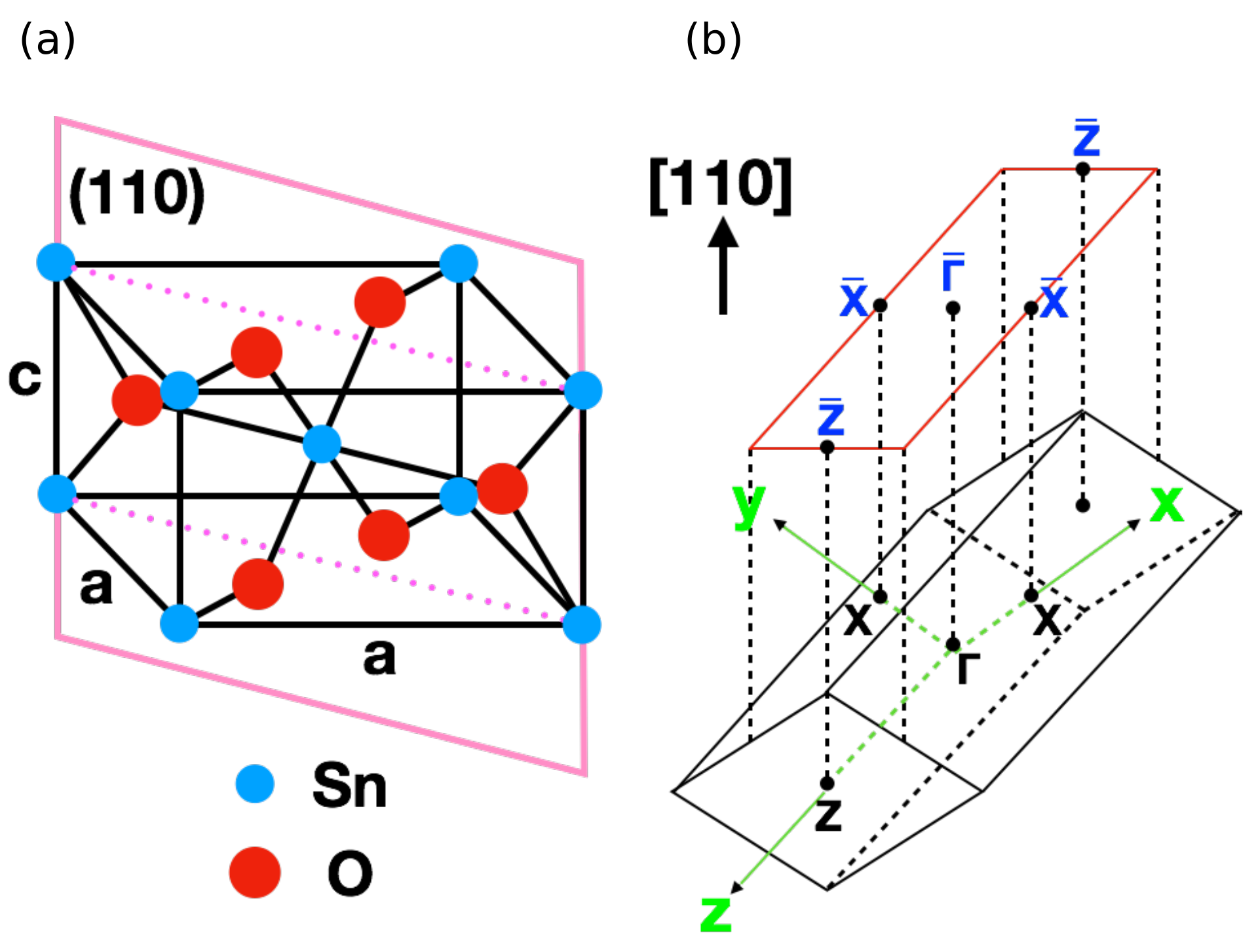} 
   \caption{
   			\footnotesize{
   			(a)~3D conventional unit cell of SnO$_2$ with $a = 4.737$~\AA~and $c = 3.185$~\AA; 
   			(b)~Corresponding 3D Brillouin zone of rutile SnO$_2$ (black parallelepiped) 
   			together with the surface Brillouin zone 
   			projected on the $(110)$ plane (red rectangle).
   			}
   			}
   \label{cry_BZ}
\end{figure}

SnO$_2$ crystallizes in the rutile structure. Similar to the perovskite lattice,
the oxygen anions form octahedra around the Sn$^{4+}$ cation. 
The primitive unit cell, shown in Fig.~\ref{cry_BZ}(a), is body-centered tetragonal. 
The corresponding 3D Brillouin zone is shown in Fig.~\ref{cry_BZ}(b), 
alongside with its projection on the $(110)$ plane. 
As will be shown later, the periodicity of the shallow metallic state 
observed in our ARPES data is in perfect agreement with the surface-projected Brillouin zone, 
confirming the 2D character of such electronic state.

\subsection{ARPES experiments and surface preparation}
\begin{figure}[h]
   \centering
   \includegraphics[clip, width=\linewidth]{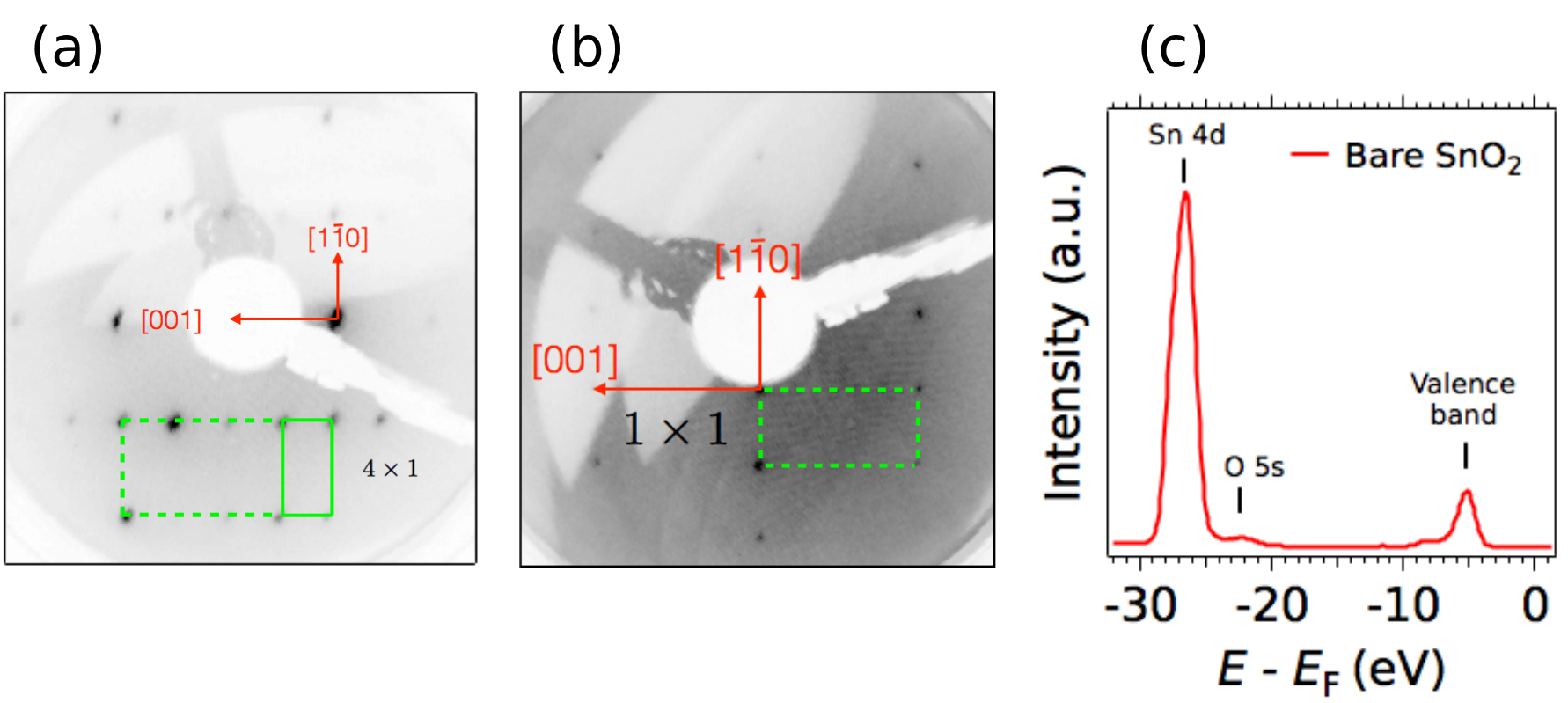} 
   \caption{
   			\footnotesize{
   			(a) LEED pattern of the bare $(110)$ surface of SnO$_2$ 
   			showing a $4 \times 1$ surface reconstruction after annealing up to 600$^{\circ}$C 
   			in UHV (electron energy: 58~eV). 
   			(b) LEED pattern of the bare $(110)$ surface of SnO$_2$ 
   			showing a $1 \times 1$ surface reconstruction after annealing up to 730$^{\circ}$C 
   			in UHV (electron energy:  116~eV). 
   			(c) Angle-integrated spectrum of the $4 \times 1$ reconstructed bare $(110)$ surface 
   			of SnO$_2$ showing the profiles of the Sn 4$d$ core-level peak and the valence band. 
   			The photon energy was 110~eV and the polarization linear horizontal. 
   			}
   			}
   \label{LEED_XPS}
\end{figure}

ARPES experiments were performed at the CASSIOPEE beamline of Synchrotron SOLEIL (France) 
and at beamline 2A of KEK-Photon Factory (KEK-PF, Japan) 
using hemispherical electron analyzers with vertical and horizontal slits, respectively. 
Typical electron energy and angular resolutions were 15~meV and 0.25$^{\circ}$. 
In order to generate pristine surfaces for the ARPES experiments, 
commercial $(110)$-oriented SnO$_2$ single crystals (SurfaceNet) were annealed for 20 minutes 
at a minimum temperature of 600$^{\circ}$C in UHV conditions. 
Low-energy electron diffraction (LEED) and core-level photoemission spectroscopy 
were employed to verify the long-range crystallinity and cleanliness of our surfaces 
after preparation. Depending on the annealing temperature, the clean surfaces 
showed either a $1 \times 1$ bulk-like periodicity
or a weak $4 \times 1$ surface reconstruction~\cite{Cox1989, Batzill2003, Batzill2005}
(see Fig.~\ref{LEED_XPS}),  
with no observable difference in the ARPES spectra of the 2DES. 
As shown as in Fig.~\ref{LEED_XPS}(c), the energy position and lineshape of photoemission peaks 
match well with previous experimental results~\cite{Batzill2005}. 

Al and Eu deposition were performed by means of molecular beam epitaxy 
using Knudsen cells, as described in Refs.~\onlinecite{Rodel2016,Lomker2017}. 
The deposition rates were calibrated with a quartz microbalance. 
Unless stated differently, ARPES data shown in this manuscript were acquired at $T \approx 16$K. 
The typical pressure during ARPES measurements was in the range of $10^{-11}$~mbar, 
while at no stage of surface preparation it exceeded $5 \times 10^{-9}$~mbar.

\section{Results}
\subsection{Creation of the 2DES}
\begin{figure}[h]
        \includegraphics[width=\linewidth]{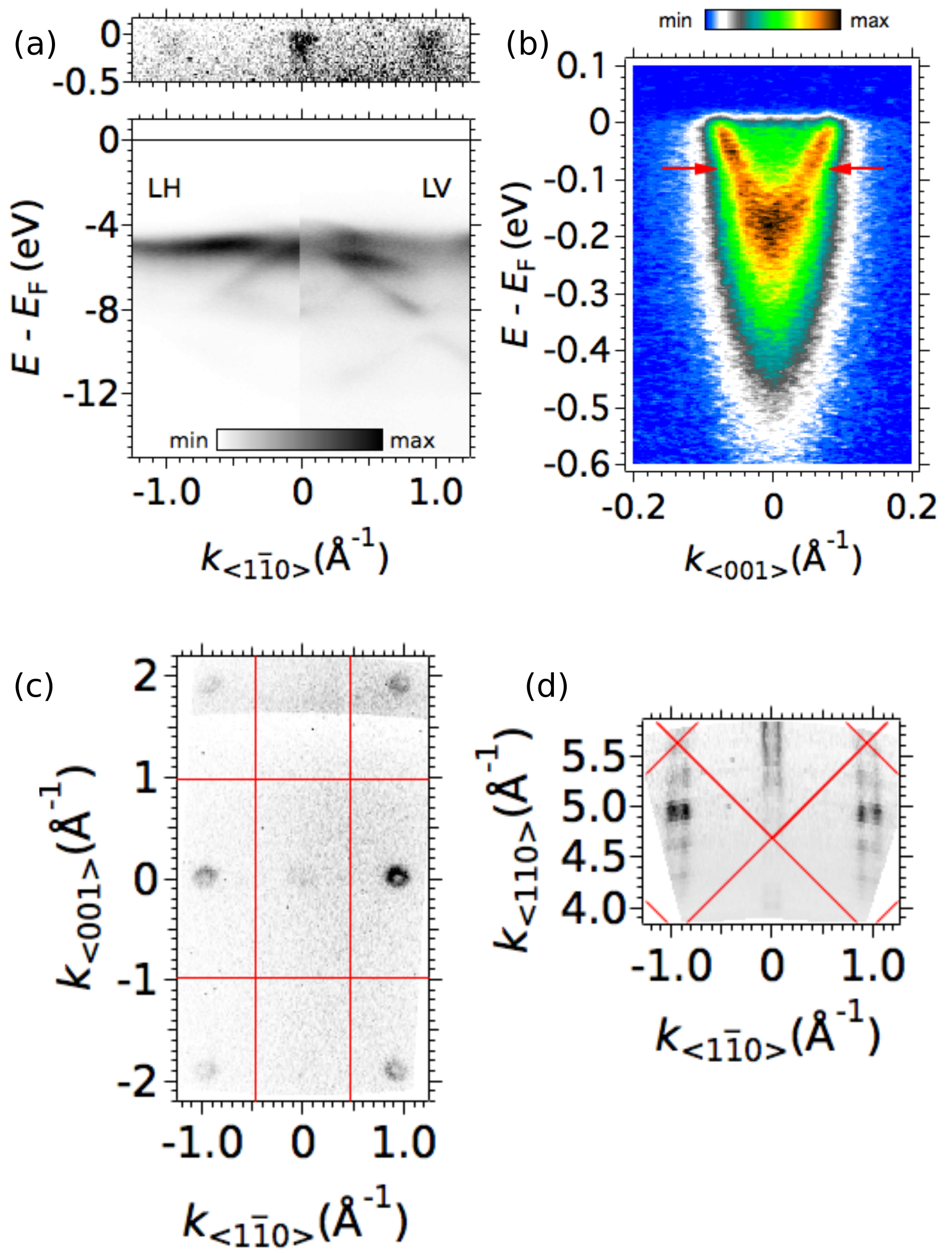}
    \caption{\label{VB_Ef}{
    		 \footnotesize{(Color online)
        	 (a) ARPES energy-momentum map of the valence band 
        	 	 on the bare $(110)$ surface of SnO$_{2}$ measured
        	 	 with photons of energy 111~eV using linear horizontal (LH, left part)
        	 	 and linear vertical (LV, right part) polarizations. 
        	 	 The top panel shows a shallow metallic state, 
        	 	 fingerprint of a 2DES, made visible after changing the contrast of the color scale. 
        	 (b) Zoom over the energy-momentum dispersion of the 2DES ($h\nu = 88$~eV). 
        	 	 Electron-phonon interaction induces the observed deviations (red arrows)
        	 	 from a simple parabolic shape. 
        	 (c) In-plane Fermi surface map of SnO$_2$(110) measured with LH 
        	 	 photons at $h\nu=88$~eV. 
        	 (d) Out-of-plane Fermi surface map of SnO$_2$(110) acquired
        	 	 by a stepwise change of the photon energy between 50 and 120~eV using LH polarization. 
        	 	 An inner potential of 10~eV was used in the calculation  
        	 	 of the out-of-plane momentum. 
        	 	 The blue dashed line indicates the $h\nu=88$~eV constant energy line, 
        	 	 where the in-plane Fermi surface in (c) was measured.
        	 	 Red lines mark the borders of the projected surface Brillouin zone in (c) 
        	 	 and of the bulk Brillouin zones in (d). 
        	 	 All data in this figure were acquired at 16 K. 
        	 }
        }
      }
\end{figure}
\begin{figure*}
        \includegraphics[clip, width=\textwidth]{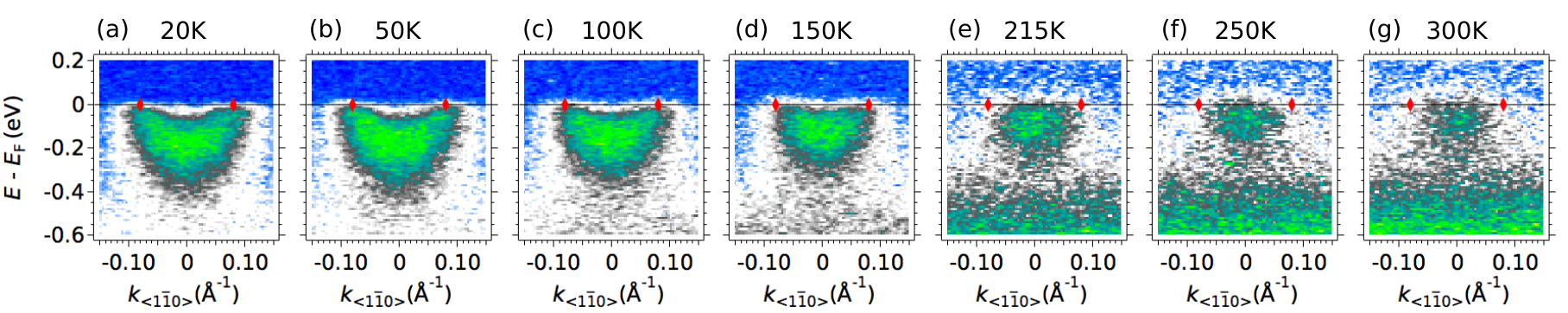}
    \caption{\label{temp_d}{
    		 \footnotesize{(Color online)
        	 Energy-momentum dispersion of the 2DES on SnO$_2$(110) 
        	 during a gradual increase of the temperature from 20~K (a) to 300~K (g). 
        	 Red markers indicate the Fermi momenta at 20K inferred from the MDC at Fermi level. 
        	 As the temperature increases, the 2DES state moves up in energy, 
        	 showing a gradual reduction of the Fermi momenta and the carrier density. 
        	 Data were acquired with $h \nu=88$~eV and LH polarisation. 
        	 }
        }
      }
\end{figure*}

Figure~\ref{VB_Ef}(a), bottom panel, shows the energy-momentum map of the valence band 
using linear horizontal and linear vertical light polarizations. 
The observed band dispersions demonstrate the crystallinity and cleanliness of the measured surface, 
while the strong dependence of the spectra on the light polarization 
indicates a well-defined orbital character of the corresponding states.
Furthermore, in striking contrast with the bulk insulating character of SnO$_2$, 
here we observe a clear metallic state in the vicinity of the Fermi level ($E_{F}$), 
with a band bottom at an approximate binding energy 
of 0.2~eV --Fig.~\ref{VB_Ef}(a), top panel. 
The energy difference with the valence band maximum (VBM) 
matches well the energy gap of SnO$_2$ (3.6~eV) as determined by previous experimental 
and theoretical studies~\cite{haines1997x, reimann1998experimental, Robertson1979, Munnix1983}. 
We therefore attribute the metallic state to the conduction band of SnO$_2$ 
that has been pulled below $E_{F}$ due to band bending, 
as observed in surfaces of other bulk insulating oxides~\cite{Santander2011, Meevasana2011, Santander2012, 
King2012, Wang2014, Rodel2014, Walker2014, Walker2015, Rodel2015, Rodel2016, Rodel2017, 
Rodel2018, Frantzeskakis2017}. However, in contrast to the $d$-orbital character 
of the conduction band in transition metal oxides, 
the conduction band bottom in SnO$_2$ originates mainly from the 5$s$ states of Sn~\cite{Robertson1979}.

Fig.~\ref{VB_Ef}(b) presents a zoom over the metallic state
with higher energy resolution. The dispersion is quasi-parabolic,
with a kink at an energy of about 80~meV (red arrows)
and an intensity built-up at the bottom of the band. 
Such a kink and strong renormalisation of the band bottom are characteristic signatures
of electron-phonon interaction, as previously observed in other oxides 
such as SrTiO$_3$ \cite{Meevasana2011,Chen2015, Wang2016}, TiO$_2$~\cite{Rodel2016} 
and ZnO \cite{Rodel2018}. 
The Fermi momentum ($k_F$) of the metallic state, 
determined from the maxima of the momentum distribution curve (MDC) 
integrated over $E_F \pm 2$meV, is $k_F = (0.077 \pm 0.001)$~\AA$^{-1}$. 

Fig.~\ref{VB_Ef}(c) shows the in-plane ARPES intensity map at $E_F$ measured over several
neighboring Brillouin zones. 
The aforementioned metallic state gives rise to a circular Fermi contour 
around the center of, and with the same reciprocal-space periodicity as, 
the surface-projected Brillouin zone of the bulk rutile structure of SnO$_2$
shown in Fig.~\ref{cry_BZ}(b).
Moreover, the isotropic shape of the Fermi contours, and especially the strong dependence of their ARPES
intensity on different Brillouin zones and/or different photon polarizations~\cite{Moser2017}
(Appendix~\ref{FS_LV}), are in line with the expected $s$-like orbital character 
at the bottom of SnO$_2$ conduction band, as discussed above.
We note that the 2DES is insensitive to the $4 \times 1$ surface reconstruction 
observed in some of our in-situ prepared surfaces, Fig.~\ref{LEED_XPS}(a),
as no Umklapp band structure is observed by ARPES. 
This suggests that the 2DES resides in the subsurface region 
(1-2 unit cells below the surface as analyzed later), 
in agreement with previous conclusions on the $(111)$-oriented surfaces 
of SrTiO$_3$ \cite{Rodel2014} and KTaO$_3$ \cite{Bareille2014}.

Fig.~\ref{VB_Ef}(d) shows the out-of-plane Fermi contour of the metallic state. 
The absence of dispersion along the $<110>$ direction throughout the complete bulk Brillouin zone 
demonstrates the near-surface confinement of the conduction band, hence forming a 2DES. 
The ARPES intensity modulations as a function of out-of-plane momentum,
also observed in many other 2DESs~\cite{Rodel2014,Rodel2015,Rodel2016,Rodel2017,Rodel2018},
are a consequence of dipole-transition selection rules 
from the confined electronic states at the surface of the material~\cite{Rodel2014,Rodel2015,Moser2018}.
The periodicity of this modulation is approximately determined by the width 
$L = 18$~\AA~of the potential well confining the electrons in the 2DES, 
which yields $2\pi/L \approx 0.35$~\AA, in good agreement with the value inferred 
from Fig.~\ref{VB_Ef}(d) --see Appendix~\ref{qw_modeling} for details.
Recalling that the in-plane Fermi surface is a circle, 
the corresponding Fermi surface in 3D $k$-space would be a cylinder. 
Our combined findings prove thus the realization of an $s$-derived 2DES on a rutile structure oxide. 
Following the Luttinger theorem, the corresponding 2D carrier density ($n_{2D}$) 
can be extracted from the area ($A_F$) of the in-plane contour as 
$n_{2D}=A_{F}/(2\pi^2) = (9.4 \pm 0.2) \times 10^{12}$~cm$^{-2}$. 
This is about 5 times smaller than the carrier density found in analogous 2DESs 
created by oxygen vacancies at the surface of other binary oxides, 
such as TiO$_2$(001)-anatase~\cite{Rodel2015} or ZnO$(000\bar{1})$~\cite{Rodel2018}, 
and up to 20 times smaller than the density of the 2DES 
at the SrTiO$_3$(001) surface~\cite{Santander2011,Meevasana2011,Rodel2016}. 

\subsection{Tunability of the 2DES}

\subsubsection{Temperature dependence}
We now show two simple methods for tuning the carrier density of the 2DES in SnO$_2$: 
controlled temperature variations and surface deposition of Europium (Eu) or Aluminium (Al). 

Figs.~\ref{temp_d}(a-g) show the energy-momentum dispersion of the 2DES 
as the temperature gradually increases from 20~K to 300~K. 
The succession of images reveals that the metallic state 
continuously shifts upwards in energy upon increasing $T$. 
The corresponding decrease of the Fermi momenta 
translates into a reduction of the 2D carrier density. 
A quantitative analysis of the $k_{F}$ values obtained from the MDCs,
presented in Tab.~\ref{kF_n2d}, shows a 50\% reduction of the carrier density at 150~K with respect to 20~K. 
We highlight the clear 2DES fingerprint at room temperature, 
but we note that there is an uncertainty whether the electronic band 
lies above or below $E_{F}$ at such high temperatures because the band bottom 
is masked by the thermal broadening. 
Temperature-dependent transport measurements may shed light on this issue. 
Nevertheless, the robustness of the 2DES at high temperatures is of obvious importance 
for technological applications. 

\begin{figure}[h] 
   \centering
   \includegraphics[width=\linewidth]{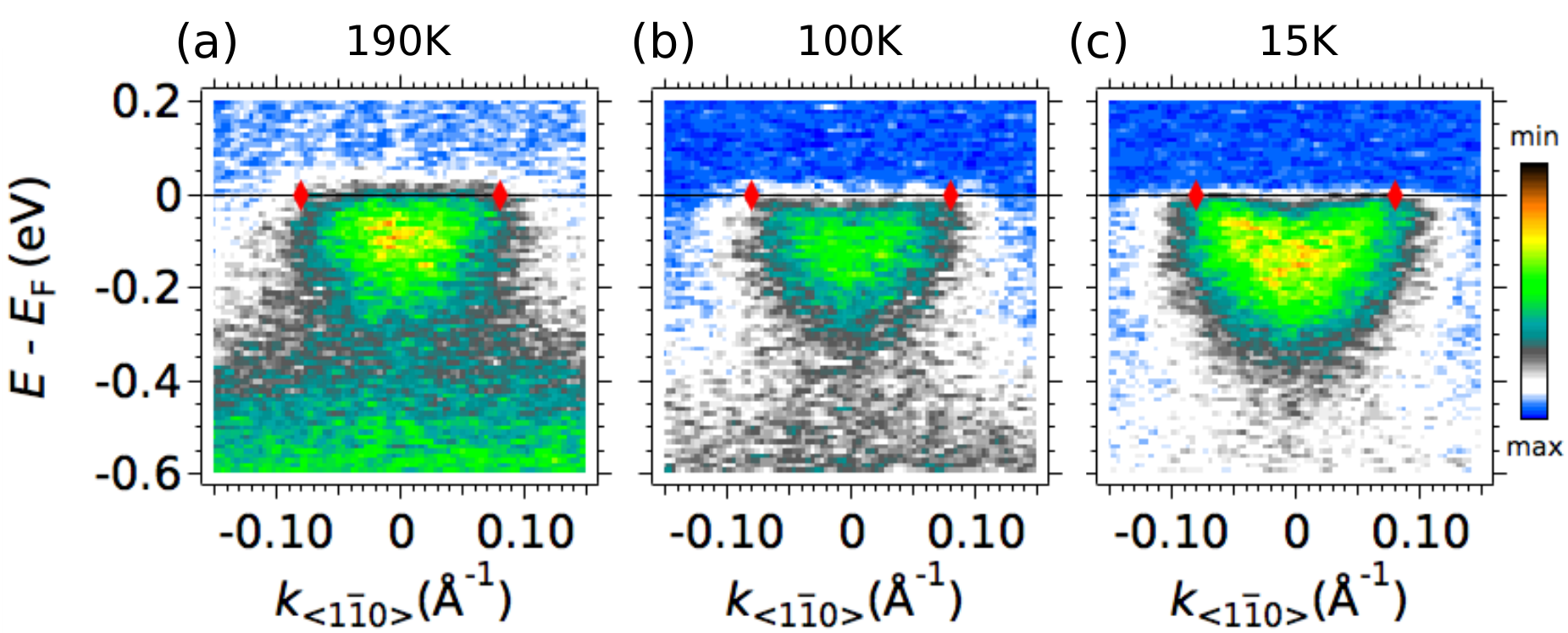} 
   \caption{
   			\footnotesize{
   			Energy-momentum dispersion of the 2DES at different temperatures 
   			decreasing from 193~K (a) to 15~K (c). 
   			As in Fig.~\ref{temp_d}, red markers denote the Fermi wave vectors 
   			of the 2DES at 20~K, before the beginning of annealing. 
   			The temperature uncertainty is $\pm~2$K.
   			}
   			}
   \label{temp_decrease}
\end{figure}

\begin{table}[h]
	\centering
	\resizebox{\linewidth}{!}{
	\begin{tabular}{c|c|c|c|c|c|c}
	\hline
	$T$(K) & 20 & 50 &  100 & 150 & $100^*$ & $15^*$  \\
	\hline
	$k_F$($10^{-3}$~\AA$^{-1}$) & 85 $\pm$ 1 & 80 $\pm$ 1 & 71 $\pm$ 1 & 60 $\pm$ 2 & 51 $\pm$ 2 & 63 $\pm$ 1 \\
	\hline
	$n_{2D}$($10^{12}$~cm$^{-2}$) & 11.5 $\pm$ 0.3 & 10.2 $\pm$ 0.3  & 8.0 $\pm$ 0.2 & 5.7 $\pm$ 0.4 & 
	4.1 $\pm$ 0.3 & 6.3 $\pm$ 0.2 \\
	\hline
	\end{tabular}
	}
	\caption{
			\footnotesize{
			Fermi wave vectors of the 2DES in SnO$_2$(110) at different temperatures. 
			The corresponding carrier densities are calculated using the Luttinger theorem. 
			A star denotes that the corresponding temperature has been reached 
			starting from a higher value (i.e, during a cooling process). 
			The temperature uncertainty is $\pm~2$K.
			}
			}
	\label{kF_n2d}
\end{table}

\begin{figure*}
        \includegraphics[width=0.9\textwidth]{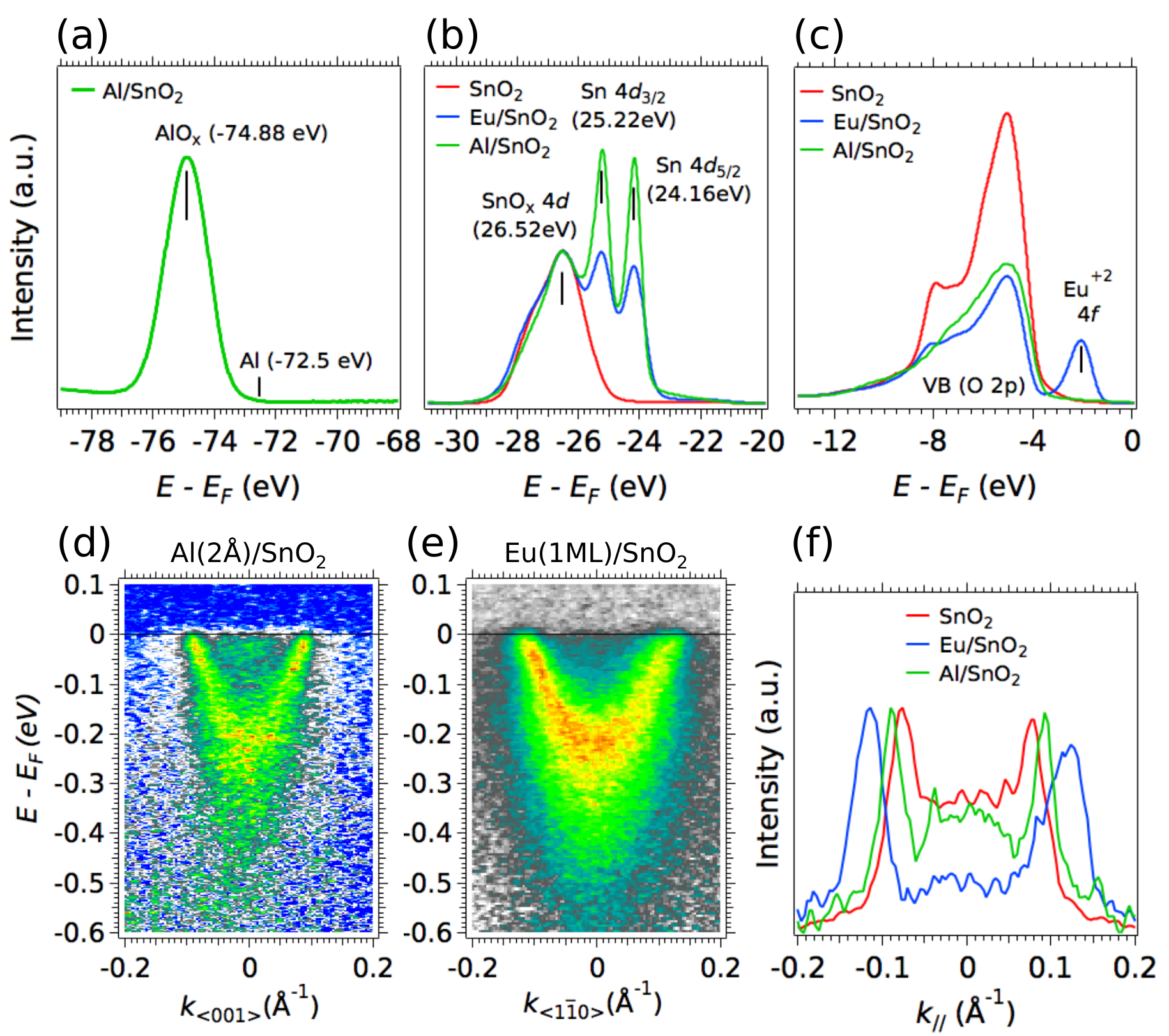}
    \caption{\label{Eu_SnO2}{
    		 \footnotesize{(Color online)
        	 (a)~Al $2p$ core-level peaks, (b)~Sn $4d$ core-level peaks, and (c)~Sn valence band 
        	 	 before (red) and after the deposition of 1~ML~Eu (blue) or 2~\AA~Al (green) 
        	 	 on SnO$_2$(110).
	 		 (d,~e)~Energy-momentum dispersion of the 2DES measured at the AlO$_x$/SnO$_2$(110)
        	 	 interface obtained after deposition of 2~\AA of elemental Al, and at the EuO/SnO$_2$(110)
        	 	 interface obtained after deposition of 1~ML of elemental Eu, respectively.
        	 (f)~Comparison of the momentum distribution curves (MDCs) at $E_F$
        	     (integrated over $\pm 10$~meV) of the 2DESs at the bare (red), Eu-capped (blue) 
        	     and Al-capped (green) SnO$_2$ surfaces.
        	 All data in this figure were measured at 16~K with LH photons. 
        	 The photon energy was 110~eV for panels (a,~b), and 88~eV for panels (c-f).
        	 }
        }
      }
\end{figure*}

We underline that the observed tunability in carrier density by means of temperature variations 
is well controlled and, to a great extent, reversible.

After heating the sample from 16~K to 300~K in order to decrease the 2D carrier density 
as shown in Fig.~\ref{temp_d}, we decreased the temperature back to 15~K 
in order to check the reversibility of the phenomenon. 
Fig.~\ref{temp_decrease} and Tab.~\ref{kF_n2d} show indeed that the 2D carrier density increases 
as the temperature decreases. After a full thermal cycle, the final $k_{F}$ value is close to the original one, 
although the recovery is incomplete, similar to previous observations for SrTiO$_3$~\cite{note2}. 

The observed temperature dependence of the 2DES carrier density in SnO$_2$
might be due to temperature-induced variations in its dielectric constant~\cite{Lanje2010},
which would affect the strength of the confining potential well, 
and/or to thermal-induced migration of oxygen atoms from the bulk to the surface. 
A thorough exploration of this issue is left open to future works.
In this sense, the reason behind the incomplete recovery of the 2DES after a warming-cooling cycle 
might be the diffusion, while the sample is heated, 
of oxygen atoms from the bulk to the surface that irreversibly replenish 
some of the oxygen vacancies.

\subsubsection{Surface deposition of Eu or Al}
The second method for modifying the 2D carrier density, 
namely surface deposition of Eu or Al, is inspired by our recent studies 
showing that some metallic adatoms on the surface of transition metal oxides act as efficient
reduction agents, pumping out the near-surface oxygen atoms, thus producing 2DES 
capped by a layer of the adatom native oxide~\cite{Rodel2016, Lomker2017}. 
Fig.~\ref{Eu_SnO2} summarizes all the pertinent spectral changes after the deposition
of 1~ML of elemental Eu or 2~\AA~of elemental Al. 
Besides the clear oxidation of Al as indicated in Fig.~\ref{Eu_SnO2}(a), 
or the formation of Eu$^{+2}$ as indicated in Fig.~\ref{Eu_SnO2}(c), 
one also observes, a shown in Fig.~\ref{Eu_SnO2}(b), that the bundle of SnO$_2$-4$d$ core levels
is drastically modified, developing two new peaks at its low binding energy side. 
The energy of these new peaks matches well with the $4d$ spin-orbit-split doublet of metallic Sn,  
while their seemingly ``inverted'' branching ratio indeed agrees well with
previous experiments on partially oxidized tin~\cite{DePadova1994},
all of which indicates that near-surface Sn atoms have been deprived of some of their oxygen neighbors.
A detailed quantitate analysis of these peaks by numerical fitting is given in 
Appendix~\ref{Sn_4d_core_fit}. 
Note that in Fig.~\ref{Eu_SnO2}(c), the additional peak above the valence band after Eu deposition, 
which we assign to the Eu$^{+2}$ $4f$ core level, did not appear after Al deposition as expected.  

Combining all these results, we can conclude that: (1) the new pair of peaks after Eu or Al deposition 
are not related to Eu core levels, but rather to the doublet of metallic Sn~$4d$ core levels; 
(2) the additional peak above the valence band after Eu deposition is not an in-gap state 
due to oxygen vacancies, but rather the Eu$^{+2}$ $4f$ core level.

The presence of Sn atoms with a lower oxidation number than 
in the pure surface is an indication that Al or Eu have been effective 
in removing the near-surface oxygen atoms by a redox reaction, 
hence driving the emergence of a 2DES, as shown respectively in Figs.~\ref{Eu_SnO2}(d) and (e), 
similar to the AlO$_x$/SrTiO$_3$ and EuO/SrTiO$_3$ systems~\cite{Rodel2016,Lomker2017}.


As shown in Fig.~\ref{Eu_SnO2}(f), the Fermi momenta of the 2DES at the AlO$_x$/SnO$_2$ interface (green) 
or the EuO/SnO$_2$ interface (blue) are larger than the one at the bare SnO$_2$ surface (red). 
The corresponding 2D carrier density in EuO/SnO$_2$, deduced from the observed 
Fermi momenta $k_F = (0.116 \pm 0.001)$~\AA$^{-1}$, is now $(2.15 \pm 0.04) \times 10^{13}$~cm$^{-2}$, 
implying a twofold increase with respect to the bare surface of SnO$_2$, 
while in AlO$_x$/SnO$_2$, the 2D carrier density increases around 40\% to 
$(1.32 \pm 0.03) \times 10^{13}$~cm$^{-2}$. 
This enhancement of $n_{2D}$ beyond the saturation limit presented by the bare surface 
under UV irradiation has not been observed in other 2DES created by redox reactions 
at metal-oxide interfaces~\cite{Rodel2016,Lomker2017},
and implies that the carrier density of the 2DES in SnO$_2$ can be tuned
via adatom deposition. As shown in the Appendix~\ref{FS_Eu_capping}, 
the 2D character of the metallic state at the EuO/SnO$_2$ interface
is confirmed by the absence of out-of-plane dispersion. 

One step further, 2~ML elementary Eu can be deposited at the surface of SnO$_2$ 
(Appendix~\ref{2ML_Eu_capping}). As a final result, 
a higher doublet of the metallic Sn~$4d$ peaks are observed, 
while no additional metallic states possibly from metallic Eu are detected, 
suggesting that the deposited 2~ML Eu are fully oxidized. 
However, the so-formed 2DES shows the same carrier density obtained with the 1~ML Eu capping,
a saturation phenomenon also observed in the EuO/SrTiO$_3$ system~\cite{Lomker2017}. 
Additionally, as 2~ML of EuO are ferromagnetic~\cite{Lomker2017}, 
one could expect to induce a magnetization of the underlying 2DES, 
which can be interesting for applications in spintronics.

\section{Discussion}
We now turn to the origin of the 2DES on SnO$_2$.
The widely-accepted mechanism of formation of 2DESs at the surface of transition metal oxides, 
such as SrTiO$_3$, TiO$_2$, KTaO$_3$, CaTiO$_3$ and ZnO, 
is the creation of oxygen vacancies in the near-surface region,
induced either by UV/X incoming photons, or by a redox reaction
between the oxide substrate and a reducing metallic agent 
evaporated at its surface~\cite{Aiura2002, Santander2011, Meevasana2011, Santander2012, 
King2012, Wang2014, Rodel2014, Walker2014, Walker2015, Rodel2015, Rodel2016, Rodel2017, Lomker2017,Rodel2018}.
Nevertheless, in some oxides, other microscopic phenomena, such as intrinsic electron accumulation 
and doping with hydrogen impurities may also have an important role 
for the onset of conductivity \cite{King2011}. 
In fact, recent numerical calculations for SnO$_2$ have suggested that 
even though the $(110)$ surface, which is characterized by an ordered arrangement 
of Sn$_3$O$_3$ clusters, contains a deficiency of oxygen atoms, 
it remains insulating with a small band gap open~\cite{Merte2017}. 
Our work shows that the deposition of a reducing metal agent 
changes dramatically the chemical environment of Sn atoms as some of the latter 
pass into a lower oxidation state [Fig.~\ref{Eu_SnO2}(a)]. 
Therefore, the structural model based on Sn interstitials 
cannot account for the surface 2DES. 
On the other hand, a structural model based on the presence of surface oxygen vacancies
as suggested in Refs.~\onlinecite{Jones1997} and~\onlinecite{Oviedo2002} 
might explain the absence of bonding counterparts to the Sn atoms and the enhancement of the 2DES.

The 2DES observed at the $(110)$ surface of SnO$_2$ is, to our knowledge, 
the first of its kind on a rutile structure. In fact, a previous comparison of the near-$E_{F}$ 
electronic structure of TiO$_2$-rutile and TiO$_2$-anatase has led to the conclusion 
that surface oxygen vacancies on a rutile structure create excess electrons 
that remain localized at Ti sites rather than forming a 2DES~\cite{Moser2013,Rodel2015}. 
The present work shows that such a conclusion cannot be generalized to all rutile lattices,
and other factors, such as the orbital character 
of the confined states ($d$ for TiO$_2$ vs. $s$ for SnO$_2$) ,
have to be taken into account to understand the emergence of a 2DES. 

\begin{figure}[!t]
        \includegraphics[clip, width=0.48\textwidth]{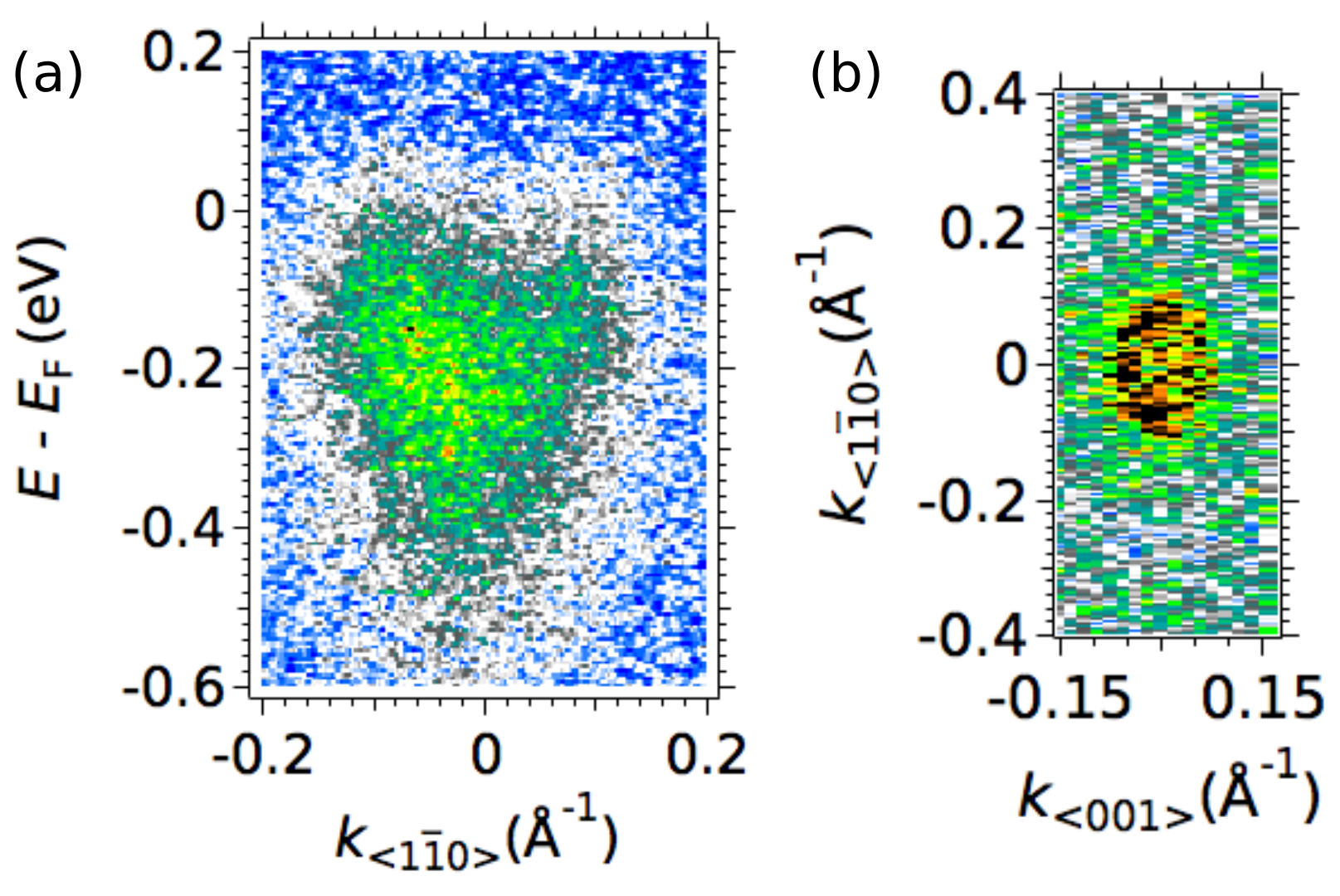}
    \caption{\label{AirExposure}{
    		 \footnotesize{(Color online)
        	 (a) Experimental energy-momentum dispersion of the 2DES 
        	 	 measured at the surface of an as-received SnO$_2$ crystal, 
        	 	 i.e., \emph{never prepared or cleaned in UHV}.
   			 (b) Corresponding Fermi surface contour around the center of the projected 
   			 	 surface Brillouin zone.
   			 All data in this figure were acquired at 20~K using LH photons of energy 120~eV.
        	 }
        }
      }
\end{figure}

Finally, we underline that besides its robustness against different surface reconstructions 
and high temperatures, the 2DES at the surface of SnO$_2$ even survives exposure to ambient pressure.
In fact, being an excellent material for heterogeneous catalysis, 
the surface of SnO$_2$ is expected to easily adsorb and desorb gas molecules. 
Thus, in order to test its catalytic properties, we studied the electronic structure of 
the SnO$_2$(110) surface \emph{without prior preparation or cleaning in UHV},
and using the UV beam at beamline 2A of KEK-PF, 
whose low brilliance ($5 \times 10^7$~photons~s$^{-1}$~$\mu$m$^{-2}$), 
about 100 times smaller than other photoemission beamlines 
in third generation synchrotrons, has been previously shown to strongly reduce or 
even inhibit the photo-induced creation of oxygen vacancies in other oxides~\cite{Backes2016}.
As demonstrated in Fig.~\ref{AirExposure},
the characteristic parabolic dispersion and circular Fermi surface
of the 2DES in SnO$_2$ can be observed even under these adverse measurement conditions
--note that even the slightest exposure to moderate vacuum would normally make the surface 
unsuitable for ARPES measurements, due to the strong surface sensitivity of the technique.  
We note furthermore that the 2DES was observed immediately after illumination
with the low-brilliance UV beam, 
and no evolution of its carrier density was evident even after several hours of measurements.

The Fermi momentum of such a 2DES intrinsically present at the SnO$_2$ surface,
determined from the MDC peaks at $E_F$, is $k_F = (0.077 \pm 0.003)$~\AA$^{-1}$, 
which is comparable to the Fermi momentum of the 2DES observed at the UHV-prepared surface 
shown in Fig.~\ref{VB_Ef}.
These results show that, either under the normal conditions of ambient pressure, 
temperature and illumination, or the exposure to low-brilliance UV light,
a 2DES in SnO$_2$ can be readily observed despite the absence of surface treatment, 
suggesting that the catalytic properties of this material protect,
or might even induce, such a 2DES,
and making of SnO$_2$ a unique candidate for future technological applications.

\section{Conclusions}
In conclusion, we observed a metallic 2DES at the $(110)$ surface of SnO$_2$ 
by means of ARPES. Its carrier density can be enhanced by the deposition of elemental Al or Eu,
and reduced by a controlled temperature increase, while the 2DES remains robust 
against different surface reconstructions, high temperatures and exposure to ambient pressure. 
The spectral fingerprints of the 2DES show an appreciable electron-phonon interaction, 
with the system always lying in the Fermi liquid regime 
(Appendix~\ref{Fermi_liquid_regime}) 
due to the small dielectric constant of SnO$_2$. 
The strong effect of a reducing metal agent such as Al or Eu on the emergence and enhancement 
of the 2DES proves that surface oxygen vacancies are at the origin of the 2DES in SnO$_2$. 
Our results provide new important insights on the long-withstanding debate 
on the origin of the $n$-type conductivity of SnO$_2$,
and prove that both the crystal structure and the orbital origin of the conduction electrons 
are decisive for the emergence of a 2DES in oxides.

\acknowledgments 
Work at CSNSM and ISMO was supported by public grants from the 
Centre National de la Recherche Scientifique (CNRS, project PICS FermiAds No 272651),
the French National Research Agency (ANR, project LACUNES No ANR-13-BS04-0006-01), 
and the ``Laboratoire d'Excellence Physique Atomes Lumi\`ere Mati\`ere'' 
(LabEx PALM projects ELECTROX, 2DEG2USE and 2DTROX) overseen by the ANR as part of the 
``Investissements d'Avenir'' program (reference: ANR-10-LABX-0039).
Work at KEK-PF was supported by Grants-in-Aid for Scientific Research 
(Nos. 16H02115 and 16KK0107) from the Japan Society for the Promotion of Science (JSPS).
Experiments at KEK-PF were performed under the approval of the 
Program Advisory Committee (proposals 2016G621, 2015S2005 and 2018S2004) 
at the Institute of Materials Structure Science at KEK.

\appendix


\section{ARPES results with linear vertical polarization}
\label{FS_LV}

\begin{figure*}
   \centering
   \includegraphics[width=0.9\textwidth]{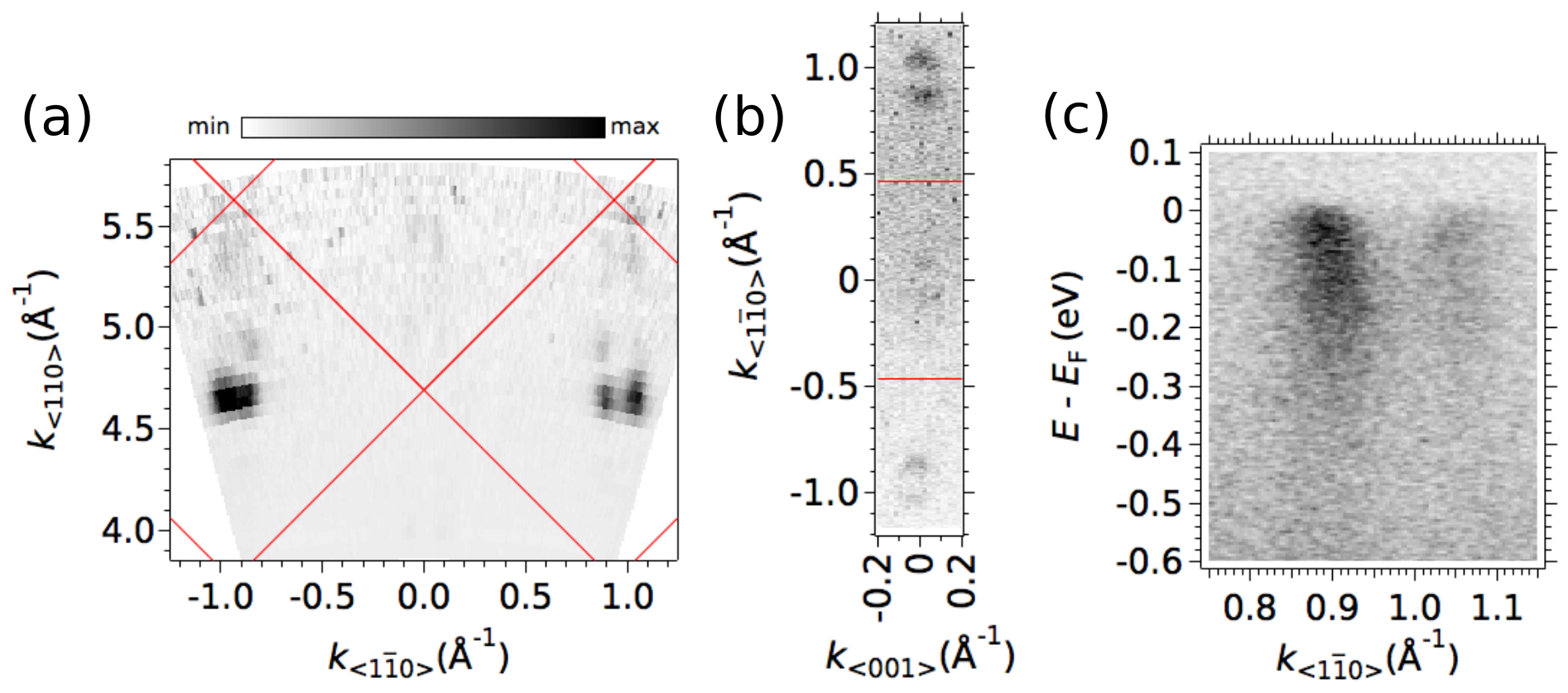} 
   \caption{
   			\footnotesize{
   			(a) Out-of-plane Fermi surface map of SnO$_2$(110) measured 
   			by changing the photon energy between 50 and 120~eV in steps of 1~eV.
   			(b) In-plane Fermi surface of SnO$_2$(110) measured with $h\nu= 88$~eV.
   			Red lines mark the borders of the projected surface Brillouin zone. 
   			(c) Energy-momentum dispersion of the 2DES measured with $h \nu = 88$~eV, 
   			corresponding to an out-of-plane momentum in the upper Brillouin zone of (b).
   			All data in this figure were acquired at 16~K using linear vertical polarization. 
   			An inner potential of 10~eV was used in the calculation of the out-of-plane momentum.
   			}
   			}
   \label{LV_all}
\end{figure*}

The majority of the data shown in the main text were acquired with linear horizontal (LH)
light polarization. For the sake of completion we present here the main results obtained 
on the bare $(110)$ surface of SnO$_2$ using linear vertical (LV) polarization. 
Fig.~\ref{LV_all} shows the out-of-plane Fermi surface contours, 
the in-plane Fermi surface contours, and the energy dispersion of the metallic 2DES 
as probed by LV photons in panels (a), (b) and (c), respectively. 
A comparison of Figs.~\ref{LV_all}(a) and Fig.~\ref{VB_Ef}(d) of the main text reveals 
a clear polarization selection at normal emission (i.e. $k_{\|}=0$) 
where there is almost no photoemission intensity with LV photons. 
Moreover, there is a strong suppression of spectral weight along the $<001>$ directions. 
These intensity variations are due to the $s$ orbital character 
of the 2DES state~\cite{Moser2017}.

\section{Gibbs free energies of relevant redox reactions}

We now consider the Gibbs free energies for the redox reactions relevant to our experiments,
namely the ones needed to form SnO$_2$, Al$_2$O$_3$, and EuO at 25 $^{\circ}$C. 
According to the database found in the chemistry-reference website~\cite{thermo}, we have:
\begin{equation}
	\begin{cases}
		\rm{Sn + O_2 = SnO_2 + 519.65~kJ/mol} \\
		\rm{Eu + (1/2)O_2 = EuO + 556.89~kJ/mol}\\
		\rm{2Al + (3/2)O_2 = Al_2O_3 + 1581.97~kJ/mol}
	\end{cases}
\end{equation}
Thus,
\begin{equation}
	\begin{cases}
		\rm{SnO_2 + 2Eu = Sn + 2EuO + 594.13~kJ/mol} \\
		\rm{3SnO_2 + 4Al = 3Sn + 2Al_2O_3 + 1604.99~kJ/mol}
	\end{cases}
\end{equation}
So, instead of a simple picture of Eu or Al ionization, these exothermic redox reactions 
will happen spontaneously once Eu or Al is deposited at the SnO$_2$ surface. 
This conclusion is consistent with our XPS data and analysis. 

\section{Quantitate analysis of Sn~$4d$ core level peaks by numerical fitting}
\label{Sn_4d_core_fit}

\begin{figure}[h] 
   \centering
   \includegraphics[width=0.9\linewidth]{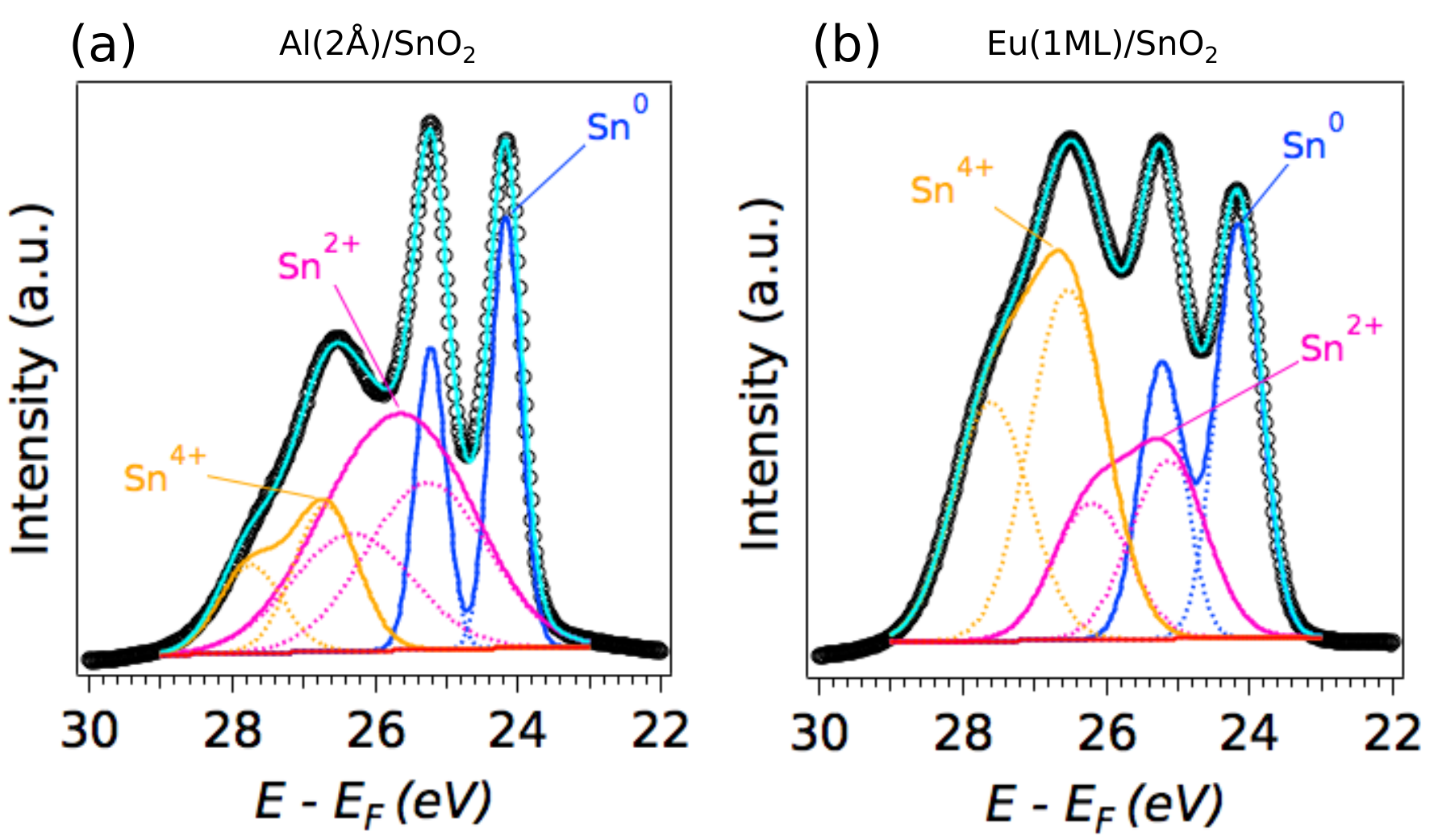} 
   \caption{\footnotesize{
   			Numerical fitting of the SnO$_2$~$4d$ core levels 
   			after deposition of (a) 2\AA~of Al, and (b) 1~ML of Eu
   			at the SnO$_2$(110) surface.
   			Black open circles: XPS data.
   			Dashed and full orange lines: Lorentzian peaks and doublet for the Sn$^{4+}$ component.
   			Dashed and full purple lines: Lorentzian peaks and doublet for the Sn$^{2+}$ component.
   			Dashed and full dark blue lines: Lorentzian peaks and doublet for the Sn$^{0}$ component.
   			Red: Shirley background.
   			Sky blue line: Overall fit.
   			}
   			}
   \label{Sn_4d_fit}
\end{figure}

We further analyzed the Sn~$4d$ core levels measured after Al(2\AA)~or Eu(1~ML) deposition 
at the SnO$_2$(110) surface by numerical fitting with Lorentzian peaks and a Shirley background 
as shown in Fig.~\ref{Sn_4d_fit}. 
As demonstrated by previous experiments~\cite{DePadova1994}, exposure of a clean tin surface to oxygen
results in the appearance of an oxidized component at the high-binding-energy side
of the metallic Sn~$4d$ doublet, composed of a bundle of four peaks:
the Sn$^{2+}$ and Sn$^{4+}$ doublets.
Thus, to fit our data we use three doublets (Sn$^{0}$, Sn$^{2+}$ and Sn$^{4+}$ states), 
constraining their branching ratio to $0.65 \pm 0.5$, 
within 10\% of the value of $2/3$ expected for a $d$-doublet,
their energy difference to $1.1 \pm 0.05$~eV, 
corresponding to the $4d$ splitting in metallic tin~\cite{DePadova1994},
and keeping the same broadening for the peaks of each doublet.
As seen in Fig.~\ref{Sn_4d_fit}, the overall fit describes very well the data,
confirming the effective reduction, and concomitant formation of oxygen vacancies,
in the SnO$_2$ surface after Al or Eu deposition.

\section{Fermi surfaces of the 2DES at the interface EuO(1~ML)/SnO$_2$}
\label{FS_Eu_capping}

\begin{figure}[h] 
   \centering
   \includegraphics[width=0.8\linewidth]{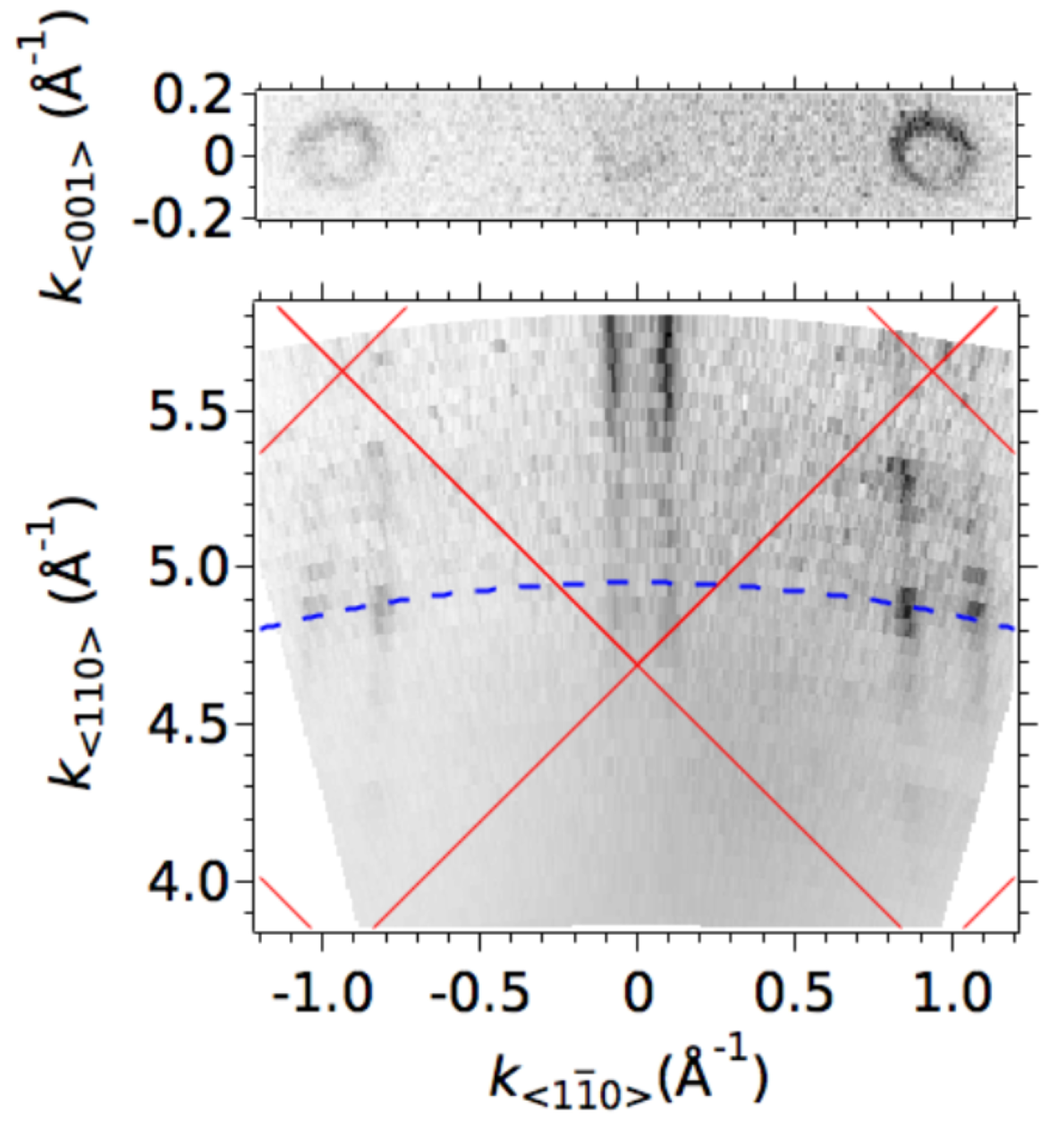} 
   \caption{\footnotesize{
   		    Top panel: in-plane Fermi surface of Eu(1~ML)/SnO$_2$(110) 
   		    measured with $h\nu = 88$~eV. 
   		    Bottom panel: out-of-plane Fermi surface map of Eu(1~ML)/SnO$_2$(110) 
   		    measured by a stepwise change of 1 eV of the photon energy between 50 and 120~eV. 
   		    The blue dashed arc shows the cut in reciprocal space corresponding to 
   		    $h\nu = 88$ eV, using an inner potential $V_0 = 10$~eV.}
   		    }
   \label{Fermi_surface_Eu_capping}
\end{figure}

Fig.~\ref{Fermi_surface_Eu_capping} shows the in-plane and out-of-plane Fermi surfaces 
obtained at the EuO(1~ML)/SnO$_2$ interface after deposition of 1~ML of metallic Eu 
on the SnO$_2$(110) surface. The in-plane periodicity of the Fermi circles 
corresponds to that of the projected bulk Brillouin zone, 
while the absence of out-of-plane dispersion demonstrates the 2D character of the confined state.

\section{2~ML elemental Eu capping at the SnO$_2$(110) surface}
\label{2ML_Eu_capping}

According to a previous work in the EuO/SrTiO$_3$ system~\cite{Lomker2017}, 
EuO(1 ~ML)/SrTiO$_3$ shows a paramagnetic behavior while EuO(2 ~ML)/SrTiO$_3$ 
shows a ferromagnetic behavior, even though the carrier density remains the same in both cases, 
indicating a saturation in the creation of itinerant electrons from oxygen vacancies at the surface. 
Inspired by this work, we deposited 2~ML pure Eu on SnO$_2$ to check whether this Eu coverage 
will also be fully oxidized, and explore potential changes in the 2DES.

\begin{figure}[h] 
    \centering
    \includegraphics[width=\linewidth]{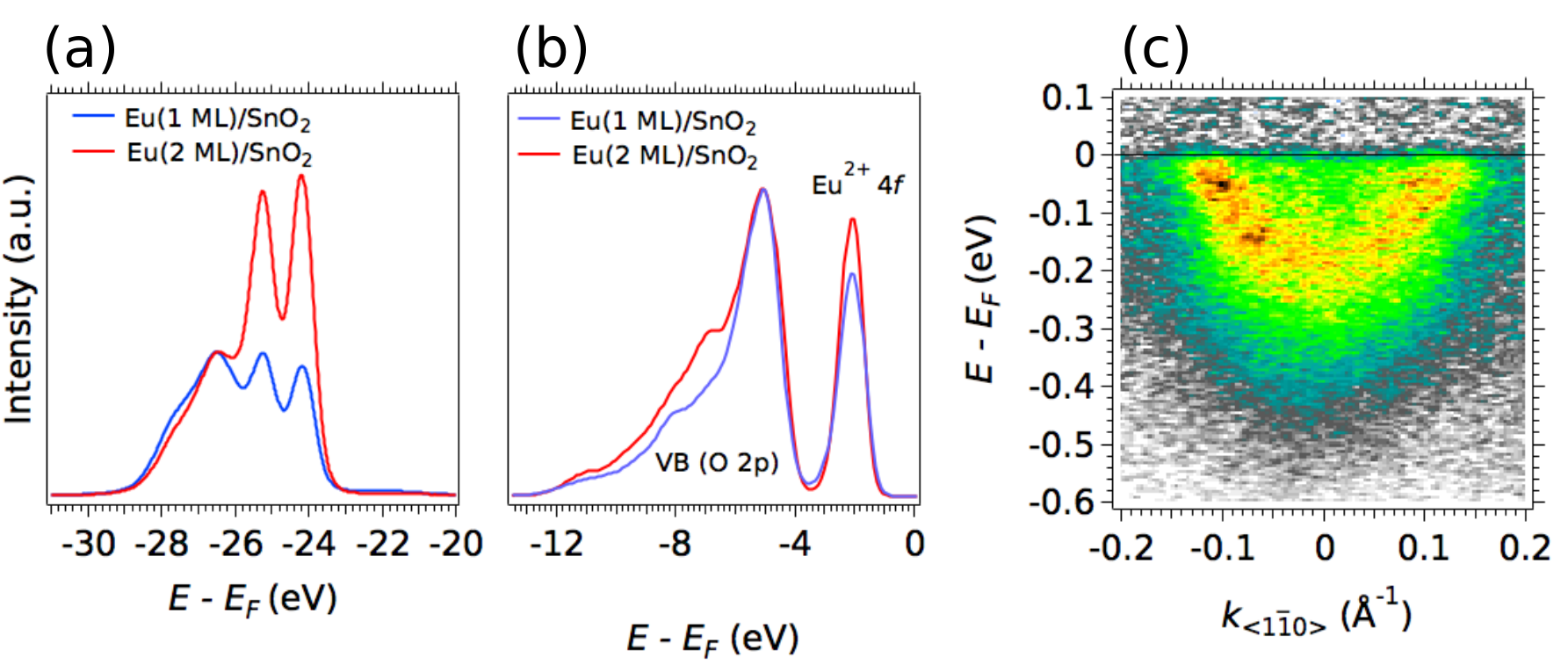} 
    \caption{\footnotesize{
    		(a, b) Sn~$4d$ core-level peaks and valence band, respectively, 
    		after the deposition of 1~ML (blue) and 2~ML (red) Eu on SnO$_2$(110). 
    		The photon energy used was 110~eV. 
    		(c) Energy-momentum dispersion of the 2DES at the Eu/SnO$_2$(110) interface 
    		obtained after deposition of 2~ML of elemental Eu measured with LH photons 
    		at $h\nu = 88$~eV.}
    }
    \label{2ML_Eu}
\end{figure}

The observation of a larger metallic Sn~$4d$ doublet in Fig.~\ref{2ML_Eu}(a) 
and a larger Eu$^{+2}$ $4f$ peak in Fig.~\ref{2ML_Eu}(b) suggest that more oxygen vacancies 
were created after deposition of 2~ML Eu on the SnO$_2$ surface, compared to 1~ML Eu deposition. 
In both cases, the XPS profile lines show the same shape without additional peaks appearing, 
indicating that Eu is still mainly oxidized into Eu$^{+2}$. 
The 2DES resulting after deposition of 2~ML Eu on the SnO$_2$(110) surface, presented 
in Fig.~\ref{2ML_Eu}(c), shows the same carrier density and effective mass as the 2DES 
at  the interface of Eu(1~ML)/SnO$_2$. An analogous saturation was reported 
in the EuO/SrTiO$_3$ system~\cite{Lomker2017}. 
These result also indicate that the 2~ML Eu are fully oxidized.

\section{Extracting the 2DES parameters}
\label{qw_modeling} 

\begin{figure}[h] 
   \centering
   \includegraphics[clip, width=0.8\linewidth]{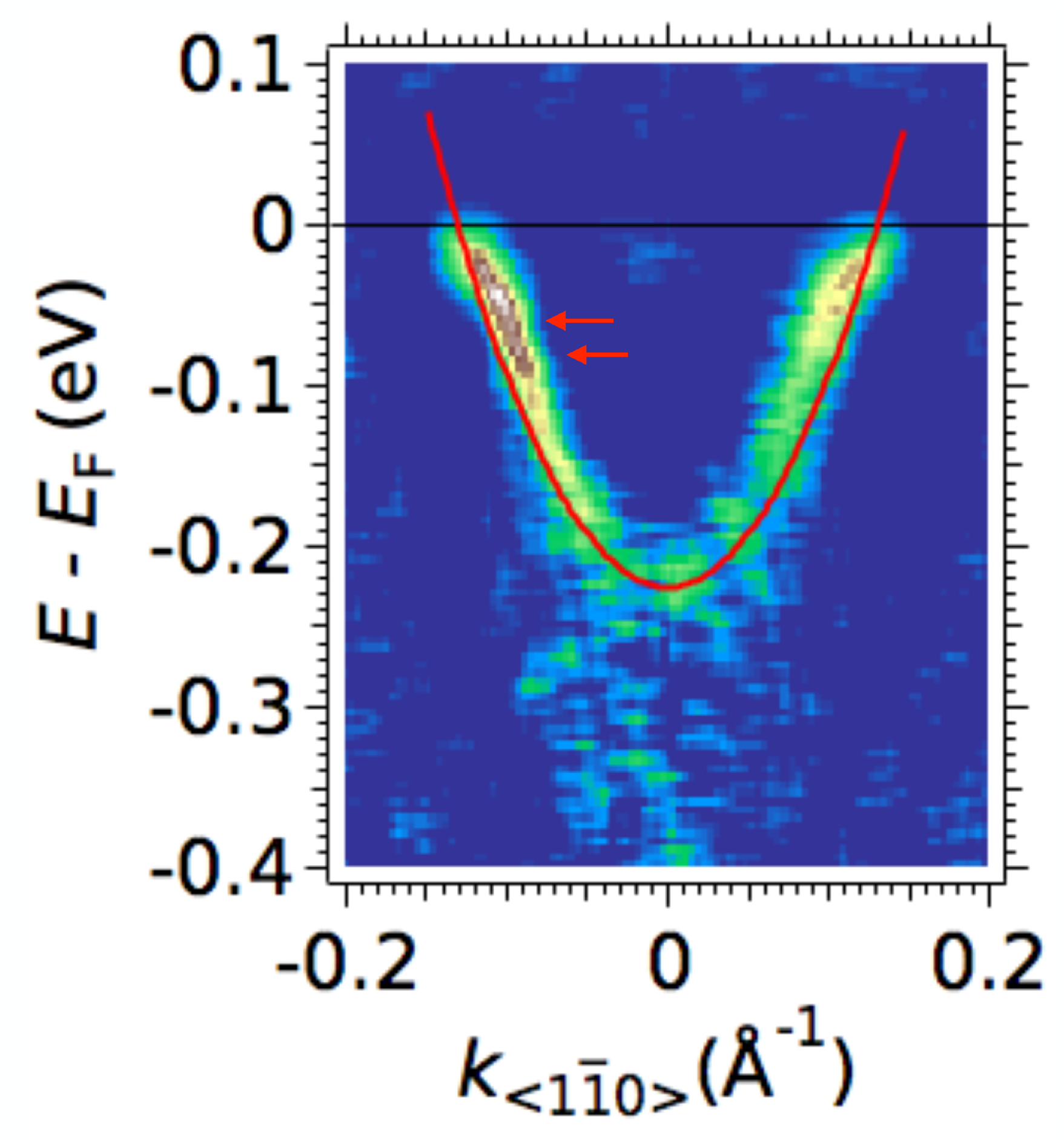} 
   \caption{
   			\footnotesize{
   			Curvature of the experimental energy-momentum dispersion 
   			shown in Fig.~3(c) of the main text together with a parabolic fit
   			of the form $E = (\hbar^2/2m^{\star})k^2 + E_{b}$,
   			with $E_b = -225$~meV and $m^{\star} = 0.29 m_{e}$.
   			Two kinks at 59~meV and 79~meV are the fingerprints of electron-phonon interaction.
   			}
   			}
   \label{2nd_curv}
\end{figure}

We now present a further analysis of the 2DES dispersion 
after the deposition of Eu in order to extract the parameters of the confinement surface potential. 
Fig.~\ref{2nd_curv} shows the 2D curvature~\cite{Zhang2011} of the spectral intensity 
previously shown in Fig.~\ref{Eu_SnO2}(e) of the main text. 
A parabolic fit of the energy dispersion, in the form $E = (\hbar^2/2m^{\star})k^2 + E_{b}$,
yields a band bottom $E_b = -225$~meV and an effective mass $m^{\star} = 0.29 m_{e}$~\cite{note3}. 
We assume that the first excited state ($n=1$) of the confining quantum well,
not directly observed in our data, lies just above $E_F$. This will provide an upper bound
for the width of the quantum well.
Taking, for simplicity, a wedge-shaped potential well, 
the values (in eV) of the discrete energy levels $E_n$ 
are given by~\cite{Santander2011, Frantzeskakis2017}:
\begin{equation*}
	E_n = V_0 + 9\times10^{-7} (\frac{m_e}{m^{\star}})^{1/3}(n+\frac{3}{4})^{2/3} F^{2/3},
\end{equation*}
with $V_0$ the energy depth of the quantum well (eV), 
$m^{\star}$ the effective mass of the confined electrons, 
and $F$ the strength of the electric field generating the confining potential (V/m).

With the $n=0$ subband (ground state) at 225~meV below $E_F$, 
and the $n=1$ subband (first excited state) at the Fermi level, 
the electric field strength can be calculated to be $F = 122$~MV/m.
This yields $V_0 \approx 500$~meV, 
$L \approx 18$~\AA~(i.e. 3 unit cells) for the spatial extension
of the wave-function in the ground state of the 2DES, the main part of which thus lies 
1-2 unit cells below the surface, and a maximum width of the confining potential 
$L_{\textmd{max}} = 42$~\AA~(6 unit cells).
These parameters are similar to the characteristics of the textbook 2DES 
observed at the $(001)$ surface of SrTiO$_3$~\cite{Santander2011, Meevasana2011, Frantzeskakis2017}.

\section{Electron-phonon interaction}
\label{Fermi_liquid_regime}

The fingerprint of the electron-phonon interaction is a kink in the experimental dispersion 
that corresponds to the characteristic energy of the phonon.
After the deposition of Eu, two such kinks are observed in the energy dispersion 
of the 2DES at the $(110)$ surface of SnO$_2$ (Fig.~\ref{2nd_curv}). 
The corresponding  phonon energies are 
59~meV ($E_g$ mode) and 79~meV ($A_{1g}$ mode)~\cite{peercy1973pressure, dieguez2001complete, 
sangeetha2011micro}. 
In the presence of electron-phonon coupling, 
the predominant interactions vary as a function of the carrier density, 
giving rise to two different regimes: the polaronic regime at low carrier densities, 
where the long-range Fr\"ohlich interaction prevails, and the Fermi liquid regime 
at high carrier densities, where the long range interaction is suppressed by 
strong electronic screening~\cite{Chen2015,Wang2016}. 

The critical carrier density between these two regimes can be estimated 
by balancing the relevant phonon energy and the surface plasma frequency $\omega_s$ 
given by the carrier density $n$~\cite{Verdi2017}:
\begin{equation*}
	\omega_s =\sqrt{ \frac{ne^2}{\varepsilon_0(\varepsilon_ {\infty}+1)m^{\star}}},
\end{equation*}
where $\varepsilon_0$ is the vacuum permittivity, 
$\varepsilon_ {\infty}$ is the dielectric constant in high frequency limit, 
and $m^{\star}$ is the carriers' effective mass. 
Given that $\varepsilon_ {\infty} = 5$ and $m^{\star} = (0.24 \pm 0.02)m_e$ for SnO$_2$, 
we can use a relevant phonon energy cutoff of 80~meV, and the aforementioned spatial extension 
of the confining potential ($L_{\textmd{max}} = 42$~\AA) 
to estimate a critical carrier density $n_{c} \approx 2.81 \times 10^{12}$~cm$^{-2}$.
This rather low critical carrier density is mainly due to the small 
dielectric constant $\varepsilon_{\infty}$ of SnO$_2$,
and means that the 2DESs both at the bare SnO$_2$ surface and at the EuO/SnO$_2$ interface
lie in the Fermi liquid regime -as their corresponding carrier concentrations
are up to one order of magnitude larger than the critical one. 
This is confirmed in the experimental data by the absence of the spectroscopic fingerprint 
of Fr\"ohlich polarons: replica bands at higher binding energies~\cite{Chen2015,Wang2016}. 
This is in stark contrast with the 2DESs on SrTiO$_3$ and TiO$_2$ surfaces 
where clear band satellites were observed at similar carrier densities 
owing to the large dielectric constant 
of the corresponding materials~\cite{Chen2015, Wang2016, Moser2013}.


\begin{thebibliography}{67}
\bibitem [{not()}]{note3}%
  \BibitemOpen
  \href@noop {} {}\bibinfo {note} {A fit of the bare data yields an identical
  energy for the band bottom but a slightly lower effective mass
  $m^{*}=0.24m_{e}$. In fact, the $k_F$ values derived from the momentum
  distribution curves of the raw data at the Fermi level are always smaller
  than the real ones because the Gaussian convolution with the experimental
  resolution shifts the band slightly towards $E_F$.}\BibitemShut {Stop}%
\end{thebibliography}

%

\end{document}